# The Lens Parallax Method:

# Determining Redshifts of Faint Blue Galaxies through Gravitational Lensing

*Matthias Bartelmann*[1,2] *and Ramesh Narayan*[1]

[1] Harvard-Smithsonian Center for Astrophysics, 60 Garden Street, Cambridge, MA 02138, USA;
[2] Max-Planck-Institut für Astrophysik, Postfach 1523, D–85740 Garching, FRG

*Abstract.* We propose a new technique, which we call the lens parallax method, to determine simultaneously the redshift distribution of the faint blue galaxies and the mass distributions of foreground clusters of galaxies. The method is based on gravitational lensing and makes use of the following: (1) the amplitude of lensing-induced distortions of background galaxies increases with redshift; (2) the surface brightnesses of galaxies decrease steeply with redshift. The distortions of galaxy images due to lensing are thus expected to be inversely correlated with surface brightness, allowing us to obtain relative distances to galaxies as a function of surface brightness. If the redshifts of the brightest galaxies are measured, then the relative distance scale can be converted to mean galaxy redshifts as a function of surface brightness. Further, by comparing the angular sizes of lensed galaxies with those of similar galaxies in empty control fields, it is possible to break the so-called mass sheet degeneracy inherent to cluster mass reconstruction techniques which are based purely on image ellipticities. This allows an unambiguous determination of the surface density of a lensing cluster. We describe an iterative algorithm based on these ideas and present numerical simulations which show that the proposed techniques are feasible with a sample of $\approx 10$ rich clusters at moderate redshifts $\sim 0.3 - 0.4$ and an equal number of control fields. The numerical tests show that the method can be used to determine the redshifts of galaxies with an accuracy of $\Delta z \approx 0.1 - 0.2$ at $z \sim 1 - 1.7$, and to measure the masses of lensing clusters to about 5% accuracy.

# 1. Introduction

Beginning with the discovery of the population of faint blue galaxies by Tyson (1988), there have been intense efforts to study the properties of faint galaxies (see Koo & Kron 1992 for a review, and references therein). The chief motivation for these studies is that faint galaxies tend to be at high redshift and thus represent early stages of galaxy evolution in the universe. Redshifts of some galaxies have been measured at the bright end of the faint galaxy population, and based on these measurements a few interesting deductions have been made (Broadhurst, Ellis, & Shanks 1988, Colless et al. 1990, 1993, Lilly, Cowie, & Gardner 1991, Koo & Kron 1992, Lilly 1993, Crampton et al. 1994, Lilly et al. 1995). However, by and large it has not been possible to obtain redshifts of the faintest galaxies. Consequently, our knowledge of the evolution of galaxies at early time in the universe is limited to what can be learned just from number counts. Unfortunately, it appears that counts alone, without redshifts, provide insufficient constraints on galaxy evolution models.

An important application of the faint blue galaxy population began with the observations of Tyson, Valdes, & Wenk (1990) who measured the distortions induced in the images of background galaxies by the gravitational lensing action of foreground clusters of galaxies. (The possibility of such distortions had been discussed in several earlier papers, e.g. Kristian & Sachs 1966, Noonan 1971, Dyer & Roeder 1976, Narayan, Blandford, & Nityananda 1984, Webster 1985, Blandford, Phinney, & Narayan 1987, Grossman & Narayan 1988.) The phenomenon is described as weak lensing and the corresponding distorted images are referred to as *arclets*, to distinguish them from the more rare strong lensing events associated with the formation of highly elongated *arcs* (see, e.g., the review by Fort & Mellier 1994, and references therein). Using the arclets in the clusters A 1689 and Cl 1409+52, Tyson et al. (1990) were able to reconstruct crude maps of the surface density distributions of the two clusters. Kochanek (1990) and Miralda-Escudé (1991) investigated how parameterized cluster mass distributions could be constrained by arclet observations. Kaiser & Squires (1993) developed this idea further and proposed a powerful technique whereby a map of the image distortions can be converted through a well-defined procedure into a two-dimensional map of the surface mass density of the lens. Their method, and variants of it (Schneider & Seitz 1995, Seitz & Schneider 1995, Schneider 1995, Kaiser et al. 1994, Kaiser 1995), are currently being used to obtain mass maps of several clusters of galaxies (Fahlman et al. 1994, Smail et al. 1995, Squires 1995). Deep observations ($B \gtrsim 26$) with present day techniques reveal $\gtrsim 10^5$ faint blue galaxies per square degree (Guhathakurta, Tyson, & Majewski 1990, Tyson 1994), so that it is possible to have as many as 2000 arclets per cluster field. It is the possibility of this huge database that makes cluster reconstruction techniques viable.

The various methods currently being used to reconstruct cluster maps suffer from two weaknesses. First, the mass estimates depend to some extent on the redshifts of the background sources. Since most of the observed arclets correspond to extremely weak background galaxies whose redshifts are not known, it is necessary to make an educated guess of the source redshifts. This leads to an unquantifiable error in the results. Secondly, as Kaiser & Squires (1993) noticed (see also Falco, Gorenstein, & Shapiro 1985), and as Schneider & Seitz (1995) have since discussed in greater detail, there is a fundamental "mass sheet" degeneracy in cluster reconstructions based on weak lensing. In brief, any reconstructed mass map can be replaced by a one-parameter family of maps, each of which is scaled by an arbitrary factor and at the same time has a mass sheet of constant mass density added to it (cf. Eq. (5.19) below); all of these models will be equally consistent with the data. The existence of the mass sheet degeneracy renders it very difficult to make quantitative statements regarding the total mass contained in a cluster.

There have been some suggestions in the literature to overcome these limitations. Smail, Ellis, & Fitchett (1994) suggested that it may be possible to obtain the redshifts of the sources directly from lensing distortions, and this idea has been discussed in further detail by Kaiser et al. (1994). Independently, Kneib et al. (1994) derived redshift estimates for a number of arclets in the field of A 370; see also Kochanek (1990). Similarly, some methods have been proposed to try and eliminate the mass sheet degeneracy in the reconstructions of the lensing clusters. The most promising of these ideas is the suggestion by Broadhurst, Taylor, & Peacock (1995) to compare the number counts of galaxies in a lensed field with the corresponding number counts in an unlensed field, thereby measuring the mean magnification induced by the cluster.

We propose in this paper a new systematic procedure, the *lens parallax method*, which may be used to obtain the redshifts of faint galaxies and at the same time to eliminate the mass sheet degeneracy. Our method makes use of two ideas. Our first idea is based on the following facts: (1) for a given lens, the amplitude of the distortions of the galaxy images induced by gravitational lensing increases with increasing source redshift in a known way; (2) the apparent surface brightnesses of galaxies decrease steeply with redshift (Koo & Kron 1992, Tyson 1994); (3) lensing leaves the surface brightness of a galaxy unchanged. Combining these, we see that lens-induced distortions are expected to increase with decreasing surface brightness. Our idea thus is that the relative distances to galaxies of different surface brightness can be inferred by studying the variation of the effects of lensing with surface brightness. Further, if the redshifts of the brightest background galaxies are measured, the relative distance scale can be calibrated, and thus the mean redshifts of the galaxies can be determined as a function of surface brightness.

Our second idea is that the magnification of galaxy images due to a lensing cluster can be inferred directly by comparing the angular sizes of galaxies in the cluster fields with the sizes of equivalent galaxies in "empty" control fields. By suitably combining this idea with the redshift determination described in the previous paragraph, we show that we can simultaneously obtain unbiased estimates of the galaxy redshifts and the surface mass densities of the lensing clusters. In particular, we show that the mass sheet degeneracy can be eliminated.

The algorithm which we propose requires three kinds of measurements: (1) The ellipticities of the faint blue galaxies in cluster fields have to be measured, together with the surface brightnesses of the galaxies. The techniques necessary for this kind of measurement have recently been considerably improved (see, e.g., Kaiser, Squires, & Broadhurst 1995, Bonnet & Mellier 1995) and successfully used to reconstruct several clusters (e.g., Fahlman et al. 1994, Smail et al. 1995, Squires 1995; see also Bonnet, Mellier, & Fort 1994). It therefore appears reasonable to expect that it will be possible in the near future to determine accurate ellipticities even of very faint background galaxies, especially in conjunction with the unprecedented resolution of the refurbished HST (as an example, see the remarkable images in Dressler et al. 1994). (2) The sizes of the faint blue galaxies have to be determined both in cluster fields and in empty control fields. This measurement should be no more difficult than the measurement of ellipticities. (3) The mean redshift of the galaxies with the largest surface brightness should be measured.

The paper is organized as follows. Sect. 2 presents the basic idea and outlines the proposed iterative algorithm in a simplified form. Sect. 3 introduces the necessary relations from gravitational lensing theory and also describes the numerical cluster models which we use in the simulations presented later. In Sect. 4, we develop a model of the background galaxies needed for the simulations. Then, in Sect. 5 we describe the algorithm in detail, presenting results from computer simulations in Sect. 6. We conclude with a summary and discussion in Sect. 7.

## 2. Outline of the method

We assume that the observations mentioned in the introduction have been performed. We are then given the following data: two-component galaxy ellipticities $\varepsilon_{1,2}(\boldsymbol{x}, S)$ and galaxy sizes $R(\boldsymbol{x}, S)$ in a number of cluster fields, and equivalent measurements, $\varepsilon'_{1,2}(S)$ and $R'(S)$, in unlensed control fields. The measurements in the cluster fields depend on the (two-dimensional) projected position $\boldsymbol{x}$ because the properties of the lenses vary spatially, whereas the data in the control fields are expected to be independent of position. All data are assumed to be given as functions of surface brightness $S$. We use the symbol $S$ to denote surface brightness in magnitudes per square arc second, and the symbol $I$ to denote surface brightness in physical units.

The number density of the faint blue galaxies is sufficiently large that even when we subdivide the cluster fields into a number of smaller patches and the surface brightness into a number of bins, we will still have a reasonable number of arclets ($\sim 10$) per patch per surface-brightness bin. From these data, we construct mean galaxy ellipticities $\langle \varepsilon(\boldsymbol{x}_j, S_i) \rangle$ and mean galaxy sizes $\langle R(\boldsymbol{x}_j, S_i) \rangle$ corresponding to the patch locations $\boldsymbol{x}_j$ and the surface-brightness bins $S_i$. Usually, the $\boldsymbol{x}_j$ would correspond to cell centers on a square grid.

In order to keep the discussion simple in this section, we assume that the lensing is weak, which means that the convergence $\kappa$ and the shear components $\gamma_{1,2}$ characterizing the local imaging properties of the lens (to be defined in Sect. 3 below) satisfy $\kappa \ll 1$, $|\gamma_{1,2}| \ll 1$. Then, the mean image ellipticities $\langle \varepsilon \rangle$ directly estimate the local shear,

$$\langle \varepsilon(\boldsymbol{x}_j, S_i) \rangle \propto \gamma(\boldsymbol{x}_j, S_i) \,, \tag{2.1}$$

where the constant of proportionality depends on the definition of the ellipticity, and where we have omitted the component index $1, 2$ on $\varepsilon$ and $\gamma$. Similarly, by comparing the mean angular sizes of the lensed galaxies with their intrinsic sizes (obtained from the control fields), we can estimate the ratio

$$r(\boldsymbol{x}_j, S_i) \equiv \left[ \frac{\langle R'(S_i) \rangle}{\langle R(\boldsymbol{x}_j, S_i) \rangle} \right]^2 = (1 - \kappa)^2 - |\gamma|^2 \approx 1 - 2\kappa(\boldsymbol{x}_j, S_i) \,. \tag{2.2}$$

The ratio $r$ is the inverse of the local magnification due to the lens; we prefer to use the inverse because it simplifies some of the relations we use later in the paper. The main point about Eq. (2.2) is that by measuring $r(\boldsymbol{x}_j, S_i)$ we can estimate the convergence $\kappa(\boldsymbol{x}_j, s_i)$, which is proportional to the local surface mass density of the lens (cf. Eq. (3.2)).

Because of the steep dependence of $S$ on the redshift $z$, we can refer to the surface-brightness bins as effective redshift bins, assigning to each surface brightness $S_i$ the average redshift $\bar{z}_i$ of the galaxies in that bin. The convergence and shear corresponding to the $i$th surface brightness bin is then related to the same quantities for the first, i.e. brightest, bin as follows,

$$(\kappa, \gamma)_i = (\kappa, \gamma)_1 d_i , \qquad (2.3)$$

where $d_i$ is a distance factor (defined in equations (5.5)–(5.8) below). Equation (2.3) simply states that the convergence and shear of galaxies with lower surface brightness are larger than those of galaxies in the brightest bin by a distance factor $d_i$ which depends only on the known redshift of the lens, and the redshifts $\bar{z}_1$ and $\bar{z}_i$ of the two surface brightness bins. We can thus estimate the distance factors from the data by means of the relation

$$d_i = \frac{\varepsilon_i}{\varepsilon_1} . \qquad (2.4)$$

There is an equivalent relation for $d_i$ based on galaxy sizes, but the ellipticity data will usually yield a more accurate result. By averaging Eq. (2.4) over all patches in a given cluster and by combining the information from several clusters (allowing for the different cluster redshifts if necessary), we can obtain an accurate estimate of the $\{d_i\}$. Combining this information with the mean redshift $\bar{z}_1$ of the galaxies in the brightest bin (which we assume to be measured), we then obtain the mean redshifts $\bar{z}_i$ of all the surface brightness $S_i$.

Simultaneously with the redshift determination, we obtain estimates of the shear field $\gamma_1$ and convergence field $\kappa_1$ corresponding to the brightest surface brightness bin as a function of position in each cluster. In general, the estimates of $\gamma_1$ will be more accurate than those of $\kappa_1$ (cf. Sect. 5). We can now proceed along one of two routes to determine the mass distribution of the cluster. First, from the map of $\gamma_1(\boldsymbol{x}_j)$, we can construct a map of $\kappa_1(\boldsymbol{x}_j)$ using one of the weak-lensing reconstruction techniques described in the literature. We use the method due to Kaiser & Squires (1993) in this paper. The resulting reconstruction of $\kappa_1$ will suffer from the mass-sheet degeneracy. However, this degeneracy can be broken using the $\kappa_1$ values which we have obtained from the galaxy magnifications, since only one choice of the mass sheet density will be consistent with the measured magnifications. The reconstructed $\kappa_1(\boldsymbol{x}_j)$ directly gives the surface density $\Sigma(\boldsymbol{x}_j)$ of the cluster (cf. Eq. (3.2)). Alternatively, we can directly take the $\kappa_1(\boldsymbol{x}_j)$ which we have obtained from Eq. (2.2) and calculate $\Sigma(\boldsymbol{x}_j)$ from it. Because of the considerable spread of intrinsic galaxy sizes compared to the relatively narrow range of intrinsic ellipticities, the second method will produce a more noisy mass reconstruction than the first. However, it has the advantage of being a very direct method, and can actually be superior for reproducing certain global parameters such as the total cluster mass.

The actual algorithm described in later sections is significantly more involved than the simple description given here because it allows for non-linearities in the lens mapping. As a result of this complication, we need to employ an iterative scheme, but the basic principle is as described here.

## 3. Gravitational lensing

We summarize in this section basic facts about gravitational lensing which we shall need later. For further details, see Schneider, Ehlers, & Falco (1992), Blandford & Narayan (1992), and Fort & Mellier (1994).

*3.1. Basics of lensing*

Gravitational lensing provides a mapping from the lens plane to the source plane, $\boldsymbol{x} \to \boldsymbol{y}$, where $\boldsymbol{x}$ and $\boldsymbol{y}$ are two-dimensional position vectors in the lens and source plane, respectively. The local properties of a gravitational lens are characterized by the Jacobian matrix $\mathcal{A}$ of the mapping,

$$\mathcal{A} \equiv \left(\frac{\partial \boldsymbol{y}}{\partial \boldsymbol{x}}\right) = \begin{pmatrix} 1 - \kappa - \gamma_1 & -\gamma_2 \\ -\gamma_2 & 1 - \kappa + \gamma_1 \end{pmatrix}, \tag{3.1}$$

where $\kappa$ is the convergence and $\gamma_{1,2}$ are the components of the shear. The convergence is given by

$$\kappa = \frac{\Sigma}{\Sigma_{\rm crit}} = \frac{4\pi G}{c^2} \frac{D_{\rm d} D_{\rm ds}}{D_{\rm s}} \Sigma, \tag{3.2}$$

where $\Sigma$ is the local surface mass density of the lens, and $D_{\rm d,s,ds}$ are the angular-diameter distances from the observer to the lens and the source, and from the lens to the source, respectively. In an Einstein-de Sitter universe, the angular-diameter distance between redshifts $z_1$ and $z_2 > z_1$ is

$$D(z_1, z_2) = \frac{2c}{H_0} \frac{1}{1+z_2} \left(\frac{1}{\sqrt{1+z_1}} - \frac{1}{\sqrt{1+z_2}}\right) \equiv \frac{2c}{H_0} \frac{1}{\zeta_2^2} \left(\frac{1}{\zeta_1} - \frac{1}{\zeta_2}\right), \tag{3.3}$$

$$\zeta_i \equiv \sqrt{1+z_i},$$

where $H_0$ is the Hubble constant.

If the intrinsic surface-brightness distribution of a source is $I'(\boldsymbol{y})$, we define its surface-brightness quadrupole as

$$Q'_{ij}(\boldsymbol{y}) \equiv \frac{\int d^2 y \, y_i y_j I'(\boldsymbol{y})}{\int d^2 y \, I'(\boldsymbol{y})}, \tag{3.4}$$

where the integrals extend over that part of the source which is enclosed by a threshold isophote and the coordinate system is centered on the image centroid. Since gravitational lensing conserves surface brightness, it follows from the definition of the Jacobian matrix $\mathcal{A}$ that the lensed image of the source has a surface-brightness quadrupole

$$Q = \mathcal{A}^{-1} Q' \mathcal{A}^{-1}. \tag{3.5}$$

Following common practice, we define the complex ellipticity, $\varepsilon = \varepsilon_1 + \mathrm{i}\,\varepsilon_2$, where

$$\varepsilon_1 \equiv \frac{Q_{11} - Q_{22}}{Q_{11} + Q_{22}}, \quad \varepsilon_2 \equiv \frac{2 Q_{12}}{Q_{11} + Q_{22}}. \tag{3.6}$$

If $\varepsilon'$ is the intrinsic ellipticity of a source galaxy, the ellipticity of its image is, according to Eq. (3.5),

$$\varepsilon = \frac{\varepsilon' + 2g + g^2 \varepsilon'^*}{1 + |g|^2 + 2\mathrm{Re}(g\varepsilon'^*)}, \qquad (3.7)$$

where the complex quantity $g$ represents a combination of convergence and shear,

$$g \equiv \frac{\gamma}{1-\kappa}, \qquad (3.8)$$

with $\gamma \equiv \gamma_1 + i\gamma_2$. Equation (3.8) shows that the ellipticities of gravitationally lensed images determine a combination of shear and convergence rather than the shear alone. In the limit of weak lensing, however, defined by $|\gamma| \ll 1$ and $\kappa \ll 1$, $g \approx \gamma$, and the image ellipticities directly provide a local estimator of the shear. This is the limit we considered in Sect. 2.

The distorting effect of the lens is usually described by means of a "distortion" parameter $\delta$ (Miralda-Escudé 1991), given by

$$\delta \equiv \frac{2g}{1+|g|^2}. \qquad (3.9)$$

If the sources are intrinsically circular ($\varepsilon' = 0$), we see from Eq. (3.7) that $\varepsilon = \delta$ and so $\delta$ is directly obtained by measuring the ellipticity of a single image. In the more general case, when the sources have intrinsic randomly-oriented ellipticities, the measured $\varepsilon$ is a combination of the intrinsic $\varepsilon'$ and $\delta$ and one must suitably average over the observed images to estimate $\delta$ (see Eq. (5.2)). To clarify our notation, the direct observables are the ellipticities $\varepsilon$ of the galaxy images. The quantities $\delta$ and $g$ describe properties of the lens, which are inferred from the $\varepsilon$ by averaging over all galaxy images within a given spatial patch and a given surface-brightness bin. Finally, the local shear $\gamma$ is related to $g$ via equation (3.8).

Equation (3.9) shows that the distortion parameter $\delta$ is invariant under the transformation $g \to 1/g^* = g/|g|^2$ (Schneider & Seitz 1995, see also Kaiser 1995), which means that a given value of $\delta$ corresponds to two solutions for $\gamma$. The physical reason for this is that there is an ambiguity regarding the parity of the image, which arises because it is not known a priori on which side of a critical curve the particular point under consideration lies. As a simple example which illustrates the situation, consider images lensed by a singular isothermal sphere. Because of circular symmetry, the lens mapping is one-dimensional, and we can consider $\gamma$, $g$, and $\delta$ as real numbers. Convergence and shear are given by

$$\kappa = \frac{1}{2x}, \quad \gamma = \frac{1}{2x}, \qquad (3.10)$$

and the tangential critical curve is at $x_\mathrm{t} = 1$, where $x$ is defined in units of the Einstein radius. Consider an image which starts at large $x$ and approaches the critical curve. As the image moves inward, $g$ and $\delta$ rise from zero towards unity. At $x = 1.5$ for instance, $g = 0.5$ and $\delta = 0.8$. Both $g$ and $\delta$ reach unity at the tangential critical curve. Proceeding inward, $\delta$ decreases and reaches the previous value of 0.8 again at $x = 0.75$, at which point $g = 2$. Comparing the points at $x = 1.5$ and at $x = 0.75$, the parity of the image has changed and so has the value of $g$. However, the image ellipticity is identical at the two points, and so the image shape does not allow us to distinguish between $g = 0.5$ and $g = 2$, i.e. between $g$ and $g/|g|^2$.

*3.2. Sample of model clusters*

In the rest of the paper we describe a detailed computer simulation of the lens parallax method. We use synthetic weak-lensing data generated by taking model cluster lenses and "observing" a population of background galaxies. Our cluster lenses are taken from a sample described and used in a series of earlier papers (Bartelmann & Weiss 1994, Bartelmann, Steinmetz, & Weiss 1995, Bartelmann 1995a,b). These clusters were produced by $N$-body simulations with $\Omega_0 = 1$, $\Lambda = 0$, $H_0 = 50$ km s$^{-1}$Mpc$^{-1}$, starting from CDM initial conditions normalized to the quadrupole moment of the microwave background fluctuations measured by COBE (e.g., Bunn, Scott, & White 1994). In total, 13 clusters were simulated, and their three-dimensional particle distributions were stored at about ten time steps per cluster between redshifts 1 and 0. The projections of these three-dimensional models along the three independent spatial directions can serve as independent cluster models for the purposes of this paper. From this sample, we have selected four clusters with redshifts between 0.35 and 0.37, whose projections provide 12 cluster lens models. The line-of-sight velocity dispersions of the clusters range over $[950, 1550]$ km s$^{-1}$, and their peak $\kappa$ values cover the range $[0.6, 1.2]$. Each cluster model is centered within a field of $5'$ side length.

In addition, we independently simulate an equal number of control fields free of lensing, from which we obtain the calibrating functions $\langle\varepsilon'(S_i)\rangle$, $\langle R'(S_i)\rangle$ for our method.

# 4. A model for the sources

We also need to generate a sample of background sources which can be lensed by the clusters. We describe the model in this section, noting that it is not our intent to account in detail for all the known features of the faint blue galaxies, but rather to come up with a reasonably realistic model which is close enough to the real situation to provide a good test of the lens parallax method.

For the purposes of the simulations, we need to specify the following intrinsic galaxy properties: redshift, luminosity, surface brightness, size, color, and ellipticity. Our choices of these parameters are guided by the following picture of the faint blue galaxies which has emerged from increasingly deep and detailed observations (see, e.g., Broadhurst, Ellis, & Shanks 1988, Colless et al. 1990, Colless et al. 1993, Lilly, Cowie, & Gardner 1991, Lilly 1993, Crampton et al. 1994; and also the review by Koo & Kron 1992). The redshift distribution of faint galaxies has been found to agree fairly well with that expected for a non-evolving comoving number density. While the galaxy number counts in blue light are substantially above an extrapolation of the local counts down to increasingly faint magnitudes, those in the red extrapolate very well. Further, while there is significant evolution of the luminosity function in the blue, in that the luminosity scale $L_*$ of a Schechter-type fit increases with redshift, the luminosity function of the galaxies in the red shows no signs of evolution. Highly resolved images of faint blue galaxies obtained from the HST are now becoming available. In red light, they reveal mostly ordinary spiral galaxies, while their substantial emission in blue light is more concentrated to either spiral arms or bulges. Spectra exhibit emission lines charateristic of star formation. These findings support the view that the galaxy evolution towards higher redshifts apparent in blue light results from enhanced star-formation activity taking place in a population of galaxies which, apart from this, basically remains the same as it is today even to redshifts of $z \gtrsim 1$.

This motivates the following choices for the model-galaxy parameters. The intrinsic redshift distribution follows from the assumption that the galaxies have a constant comoving number density. We assume that their luminosity function resembles the Schechter function of nearby galaxies. Based on the observation that the faint blue galaxies are mainly spirals, it appears reasonable to assume that they obey Freeman's law (Freeman 1970), which states that the face-on surface brightness of disk galaxies is approximately constant over a wide range of intrinsic luminosity. The size is determined once the luminosity and surface brightness are specified. We model the color distribution by a Gaussian, and adopt an ellipticity distribution which resembles the observed distribution. We elaborate on the details in the following subsections.

## 4.1. Intrinsic redshift distribution

Let $n_0$ be the constant comoving number density of galaxies. Then, in an Einstein-de Sitter model universe, there are

$$dN = 16\pi n_0 \left(\frac{c}{H_0}\right)^3 \frac{1}{(1+z)^{3/2}} \left(1 - \frac{1}{\sqrt{1+z}}\right)^2 dz \qquad (4.1)$$

galaxies within a redshift interval $dz$ of $z$. Upon substituting $\sqrt{1+z} \equiv \zeta$ as a redshift variable, the normalized redshift distribution of the galaxies can be written

$$p_z(\zeta) \propto \left(\frac{dN}{dz}\right)\left|\frac{dz}{d\zeta}\right| = 3\left(\frac{\zeta_{\max}}{\zeta_{\max}-1}\right)^3 \frac{(\zeta-1)^2}{\zeta^4} \quad , \quad \zeta \leq \zeta_{\max} , \qquad (4.2)$$

where $\zeta_{\max}$ represents a possible upper redshift cutoff of the galaxy population.

## 4.2. Luminosity, surface brightness, and color

As mentioned earlier, we assume that the luminosities of galaxies follow a Schechter-type distribution function,

$$p_L(\ell) \propto \ell^{-\beta} \exp(-\ell) \qquad (4.3)$$

(Schechter 1976), where $\ell \equiv (L/L_*)$. We take the faint-end exponent $\beta$ and the luminosity scale $L_*$ from Efstathiou, Ellis, & Peterson (1988),

$$\beta = -1.07 \quad , \quad L_* = 1.1 \times 10^{10} L_\odot , \qquad (4.4)$$

and we assume $L_*$ to be independent of redshift for simplicity; i.e., we do not include any redshift evolution of the blue luminosity. The absolute normalization of $p_L(\ell)$ is irrelevant for our purposes, because we choose the number of galaxies distributed within a cluster field so as to agree with the observed number density.

The surface brightness of spiral galaxies has been observed to be fairly constant over a rather wide range of magnitudes (Freeman's law, Freeman 1970; van der Kruit 1987,

Lauberts & Valentijn 1989). The surface-brightness distribution peaks at $S_0 \approx 21.5$ mag/arc sec$^2$, with a full width at half maximum of $\Delta S \approx 1$ mag/arc sec$^2$ and a tail towards low surface brightnesses (van der Kruit 1987, Schechter 1995, private communication). As a simple representation of such a distribution, we assume that the physical surface brightness $I$ (in physical units, not magnitudes per square arc seconds) is Gaussian distributed,

$$p_{\rm I}(I) \propto \exp\left(-\frac{(I-I_0)^2}{2\sigma_{\rm I}^2}\right) , \qquad (4.5)$$

where $\sigma_{\rm I} = 0.5\,I_0$, and $I_0$ corresponds to $S_0 = 21.5$ mag/arc sec$^2$. We neglect a possible dependence of the surface brightness on the inclination angle of disk galaxies.

The variations of the magnitude and surface brightness of a galaxy as functions of redshift depend on the color of the galaxy. We express the color in terms of a spectral index $\alpha$, such that the spectral energy distribution $f_\nu \propto \nu^\alpha$, where $\nu$ is the frequency. The $k$-correction for the galaxy magnitude is then given by

$$k = -2.5\,(1+\alpha)\log_{10}(1+z) . \qquad (4.6)$$

If $\alpha = -1$, then $\nu f_\nu$ is independent of $\nu$ and the $k$-correction is zero. In this limit, we know that the surface brightness is proportional to $(1+z)^{-4}$. For a general $\alpha$, the redshift dependence of the surface brightness is given by

$$I \propto (1+z)^{\alpha-3} \quad , \quad S \to S - 2.5(\alpha-3)\log_{10}(1+z) . \qquad (4.7)$$

Very blue galaxies have a $B-V$ color $\lesssim 0.7$, which corresponds to $\alpha = -1$ (see, e.g., Lilly, Cowie, & Gardner 1991), while "redder" galaxies have smaller values of $\alpha$. We adopt a Gaussian spectral-index distribution,

$$p_\alpha(\alpha) \propto \exp\left(-\frac{(\alpha-\alpha_0)^2}{2\sigma_\alpha^2}\right) , \qquad (4.8)$$

with $\alpha_0 = -1.5$ and $\sigma_\alpha = 0.6$. This choice appears to provide a fairly realistic distribution of galaxy colors.

### 4.3. Size and ellipticity

Having specified the surface brightness $I$ and the luminosity $L = \ell L_*$ of the galaxies, a measure for their intrinsic linear size is given by

$$R' \propto \sqrt{\frac{L}{I}} , \qquad (4.9)$$

where the unit of $R'$ is arbitrary. The angular size of a galaxy in the absence of lensing is of course $R'/D_{\rm s}$.

Assuming that the emission of the faint blue galaxies is dominated by their disk component, their intrinsic ellipticity distribution should be determined solely by their random inclination angles relative to the line-of-sight, modified perhaps by the effects of dust (see, e.g., Rix 1995 and references therein). This, however, yields an ellipticity distribution with a larger mean and variance than that observed. We have therefore chosen to use the observations as a guide and to model the intrinsic ellipticities in terms of the following two-dimensional distribution,

$$p_{\mathrm{e}}(\varepsilon_1, \varepsilon_2) = \frac{\exp\left(-|\varepsilon|^2/\sigma_\varepsilon^2\right)}{\pi\sigma_\varepsilon^2\left[1 - \exp(-1/\sigma_\varepsilon^2)\right]} , \qquad (4.10)$$

where $|\varepsilon| \equiv \sqrt{\varepsilon_1^2 + \varepsilon_2^2}$. If the image is an ellipse, $\varepsilon$ is given by

$$|\varepsilon| \equiv \frac{a^2 - b^2}{a^2 + b^2} , \qquad (4.11)$$

where $a$ and $b$ are the major and minor axes of the ellipse, respectively. We distribute the position angles of the ellipses uniformly over $[0, \pi]$. Based on the work of Miralda-Escudé (1991), Tyson & Seitzer (1988), and Brainerd, Blandford, & Smail (1995), we choose $\sigma_\varepsilon = 0.3$.

### 4.4. Detection criterion

Finally, we need a criterion to determine whether or not a particular simulated galaxy will be detected. Once we have selected all the properties described above for a galaxy, we calculate its apparent magnitude using the $k$ correction given in Eq. (4.6) and its apparent surface brightness using Eq. (4.7).

The detectability of a galaxy depends both on its apparent magnitude and surface brightness. The signal received from the galaxy is proportional to the total energy received, and thus to the flux $F$. The noise on the other hand depends on the number of pixels covered by the galaxy image,

$$\mathrm{noise} \propto \sqrt{N_{\mathrm{pixel}}} , \qquad (4.12)$$

and is therefore sensitive to the linear size of the image. Hence,

$$\mathrm{signal} \propto F , \ \mathrm{noise} \propto \sqrt{\frac{F}{I}} , \qquad (4.13)$$

and

$$\frac{\mathrm{signal}}{\mathrm{noise}} \propto \sqrt{FI} . \qquad (4.14)$$

If we assume that all galaxy images can be detected for which the signal-to-noise ratio exceeds a specified constant value, and if we further express $F$ and $I$ in terms of magnitudes $B$ and magnitudes per square arc second $S$, respectively, we obtain the criterion

$$B + S \leq \mathrm{const.} \equiv (B + S)_{\mathrm{max}} , \qquad (4.15)$$

where the constant has to be chosen appropriately. Eq. (4.15) expresses the expectation that the lower the surface brightness of a galaxy the harder it is to detect it at a given magnitude. For the detection threshold, we have chosen $(B + S)_{\mathrm{max}} = 53$.

## 4.5. Properties of the simulated source population

Figures 1–4 display some of the properties of the synthetic galaxy population specified by the model described above.

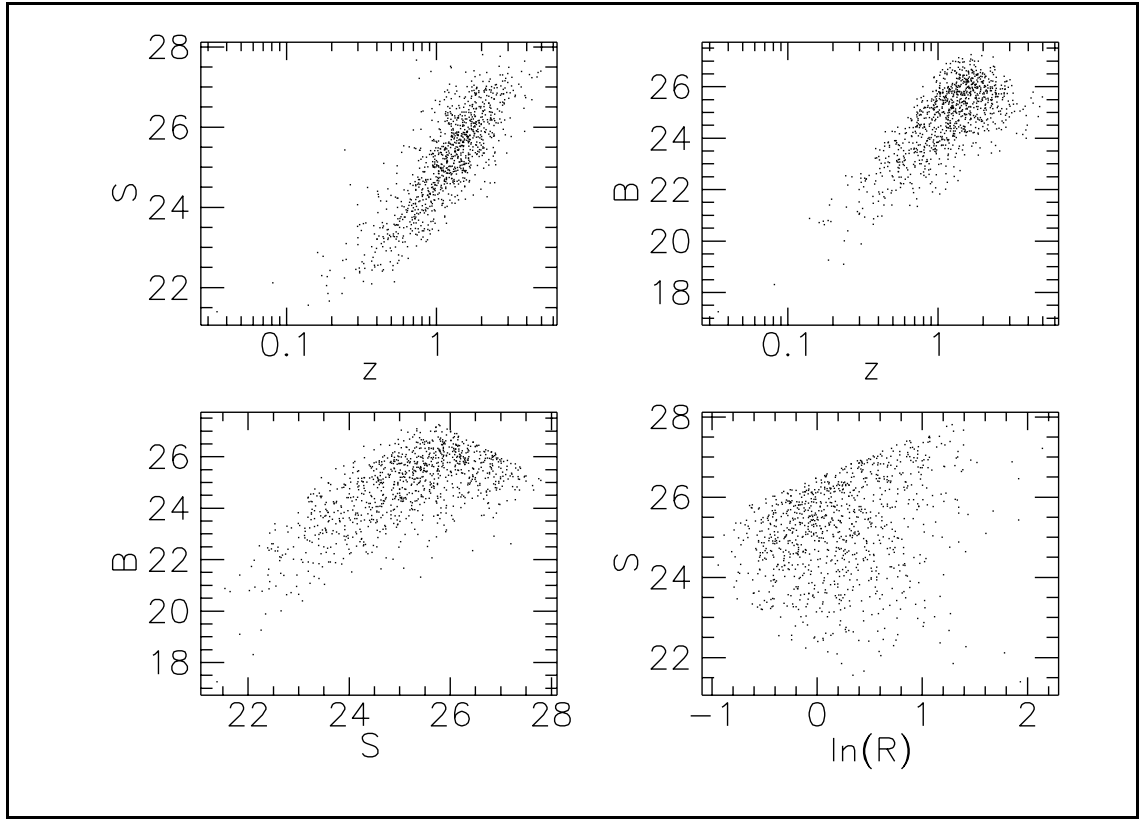

Figure 1.– Scatter plots of observables for a galaxy population whose intrinsic properties are drawn from the distributions specified in Sect. 4. Upper frames: correlations between surface brightness $S$ and redshift $z$ and between blue magnitude $B$ and $z$ (left and right frames, respectively). Note that the $S$-$z$ relation is tighter than the $B$-$z$ relation. Lower left frame: correlation between $B$ and $S$; the cutoff towards the upper right corner is due to the detection criterion, Eq. (4.15). Lower right frame: correlation between $S$ and $\ln(R)$ where $R$ is the apparent size of the galaxies; again, the cutoff towards the upper left corner is due to the detection threshold.

Figure 1 shows scatter plots of the distributions of $10^3$ galaxies in various cuts through the parameter space spanned by $(B, S, z, R)$. The intrinsic galaxy properties were drawn randomly from the distributions specified above, and the galaxies were then selected according to the criterion of Eq. (4.15), viz. that the sum of their apparent magnitude and apparent surface brightness should be smaller than $(B + S)_{\mathrm{max}}$. The upper two frames show plots of the apparent surface brightness $S$ and apparent magnitude $B$ versus galaxy redshift $z$. We see that the correlation between apparent surface brightness $S$ and redshift $z$ is fairly tight. This is partly a consequence of Freeman's law, but it is also in large part because of the steep dependence of the apparent surface brightness on redshift expressed by Eq. (4.7). The fact that $S$ changes rapidly with redshift, and is at the same time invariant to lensing, makes it a particularly favorable galaxy label for the type of investigations we propose. The lower two frames in Fig. 1 display the relation between $B$ and $S$ (left frame) and between $S$ and $\ln(R)$ (right frame). The sharp cutoffs for $S \gtrsim 26$ are caused by the detectability criterion (4.15).

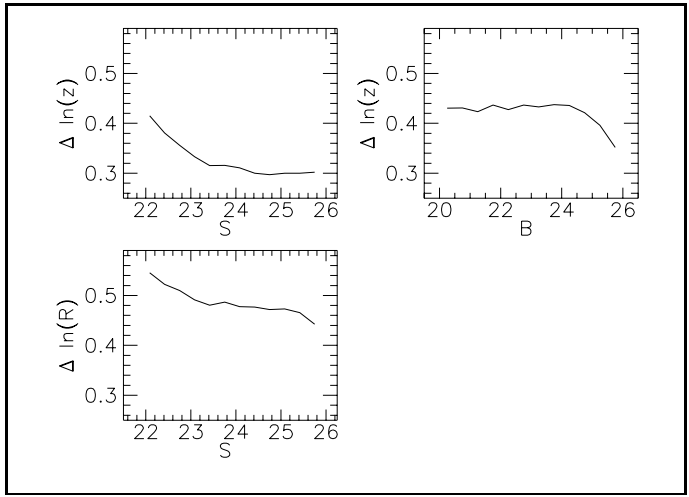

Figure 2.– Upper frames: variance $\Delta \ln(z)$ of the relations between $S$ and $z$ and between $B$ and $z$ (left and right frame, respectively), as functions of $S$ and $B$. The variance of the $B$-$z$ relation is about 50% larger than the variance of the $S$-$z$ relation. Lower frame: variance $\Delta \ln(R)$ of the logarithmic size of the galaxies as a function of $S$. $\Delta \ln(R) \approx 0.5$ at all $S$.

Figure 2 further quantifies the tightness of the correlations displayed in Fig. 1. The upper two frames show the variance of the logarithmic redshift $\Delta \ln(z)$ as a function of apparent surface brightness and magnitude. In other words, the curves in these two frames show the width of the galaxy distributions in redshift as a function of $S$ and $B$. We see that the variance of the $S$-$\ln(z)$ relation is approximately three quarters of the variance of the $B$-$\ln(z)$ relation, which quantifies the visual impression from Fig. 1 that surface brightness $S$ is a more reliable measure of redshift than magnitude. The bottom frame shows that the variance of $\ln(R)$ is approximately 0.5, quite independent of $S$. This agrees well with the scatter seen in faint galaxy images observed with HST (Kaiser 1994, private communication). The increase of $\Delta \ln(z)$ for increasing apparent surface brightness (decreasing $S$) is due to the fact that at bright $S$ the redshift is small and hence $\ln(z)$ is very sensitive to the value of $z$, while the decrease in $\Delta \ln(z)$ for increasing $B$ magnitude is a consequence of the detection threshold imposed.

Figure 3 shows the predicted differential and cumulative distribution functions of observed galaxies as a function of apparent surface brightness (upper left frame), apparent magnitude (upper right frame), redshift (lower left frame) and ellipticity (lower right frame). The solid lines show the differential distributions, and the dotted lines the cumulatives. One might expect that the differential distributions of the galaxies with $S$ and $B$ should rise monotonically when the objects become fainter. This is not the case because of the detection criterion which incorporates a cutoff in both apparent surface brightness and magnitude. If the magnitude becomes large (the galaxies become faint), the number of observable galaxies drops because of the requirement that they must then have a high surface brightness to ensure detection. The median of the redshift distribution is close to unity, but the redshift distribution has a low-amplitude tail towards high $z$, which corresponds to a tail of extremely bright galaxies with $L \gtrsim L_*$. The average apparent surface brightness is $\bar{S} \approx 25.5$ mag/arc sec$^2$, and the average apparent magnitude is $\bar{B} \approx 24.8$ .

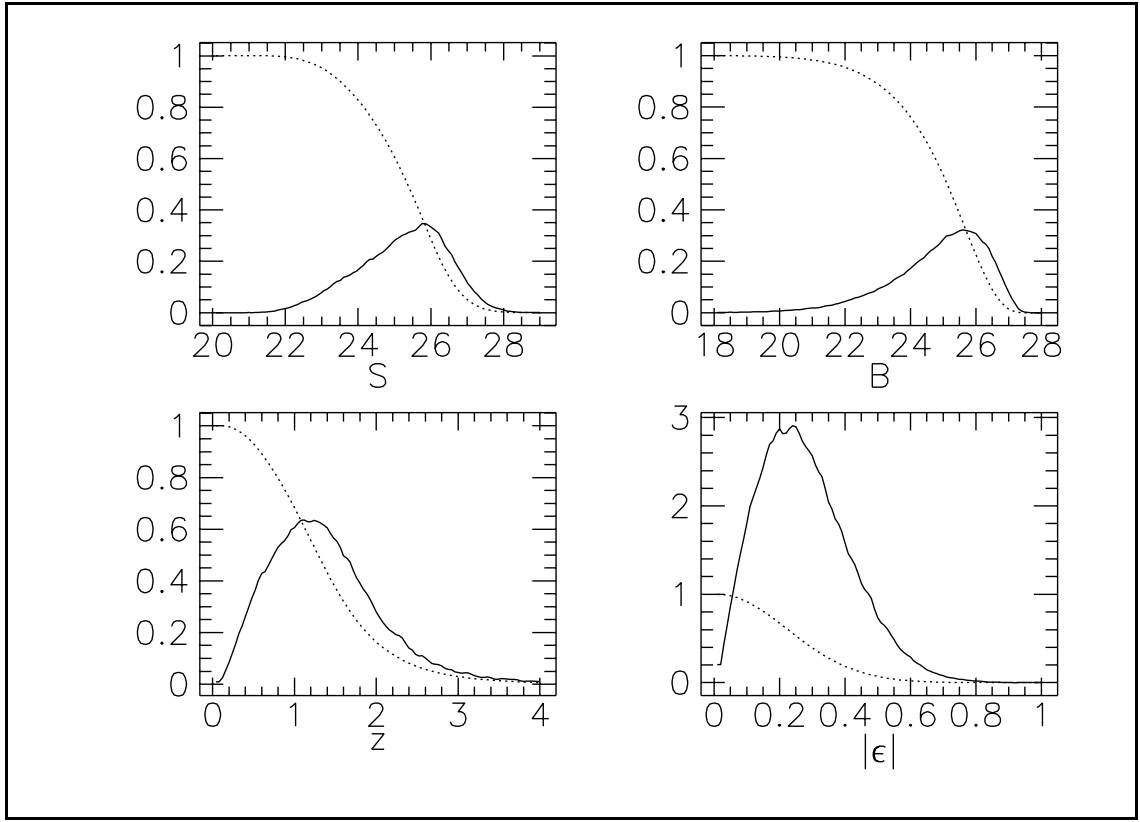

Figure 3.– Differential and cumulative distribution functions (solid and dotted lines, respectively) of observable galaxies. Upper frames: distributions in surface brightness $S$ and magnitude $B$ (left and right frames, respectively); lower frames: distributions in redshift $z$ (left) and ellipticity $\varepsilon$ (right).

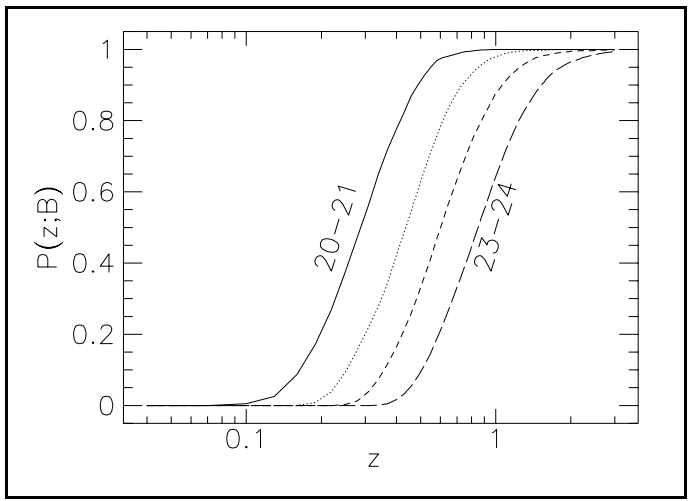

Figure 4.– Cumulative redshift distributions of observable galaxies in four adjacent magnitude bins between $B = 20$ and $B = 24$, distinguished by line type as indicated. The median redshifts in the four bins are $\bar{z} \in \{0.29, 0.44, 0.60, 0.85\}$.

Figure 4 finally shows the cumulative redshift distributions of observed galaxies in four adjacent $B$-magnitude bins, each of which is one magnitude wide. The leftmost curve is for galaxies with $B \in [20, 21]$, the rightmost for $B \in [23, 24]$. These curves are in good agreement with those given by Koo & Kron (1992, Fig. 4 and the lower panel of Fig. 5). The median redshifts of the distributions are 0.29, 0.44, 0.60, and 0.85, for the four magnitude bins given. While the median redshifts for the brighter bins agree very well with observationals, but the median redshift of the faintest bin appears too high by $\approx 0.05$.

On the whole, we feel that the model galaxy population we obtain with the various prescriptions described in this paper are reasonably close to the real galaxy population and therefore adequate for the simulations described below. Since all we are trying to do is test the method we propose, this level of realism seems adequate.

For the simulations, we populate the cluster and control fields with galaxies drawn at random and arranged to have a mean number density of $n = 70 \, (\text{arc min})^{-2}$ (Tyson 1994; Kaiser 1994, private communication). We distort the images of galaxies in the cluster fields with redshifts $z$ larger than the cluster redshifts $z_\text{d}$ according to the local convergence $\kappa$ and shear $\gamma_{1,2}$ of the lens and then calculate the observed size and distortion via Eq. (3.7). Galaxies with $z \leq z_\text{d}$ are included unperturbed. Galaxies are considered observable when their apparent $B$ magnitude, after being magnified by the lens, together with their apparent surface brightness $S$, match the detectability criterion of Eq. (4.15). Hence, for each model cluster a list of image positions, magnitudes, surface brightnesses, ellipticities, and linear sizes is created. The galaxy control fields are left unperturbed (that is, we neglect any additional weak distortion that may arise from large scale structure in the universe, cf. Blandford et al. 1991, Miralda-Escudé 1991, Kaiser 1992).

## 5. Method of analysis

We define $N_\text{bin}$ surface-brightness bins between $S_\text{min}$ and $S_\text{max}$, where $S_\text{min}$ is chosen such that galaxies with $S \sim S_\text{min}$ are at a higher redshift than the lensing cluster, and $S_\text{max}$ is selected so as to have enough galaxies in the faintest bin for a reasonable signal-to-noise ratio. In our simulations, we have chosen $S_\text{min} = 23.5$ and $S_\text{max} = 27$ (cf. the scatter plot in the upper left frame of Fig. 1), dividing this range into $N_\text{bin} = 10$ equidistant bins. For each cluster, we sort the galaxy images into these bins. Then, from the galaxy ellipticities and sizes, we determine separate distortion fields $\delta_i(\boldsymbol{x})$ and inverse magnification fields $r_i(\boldsymbol{x})$ for each surface-brightness bin $S_i$ and for each cluster.

As described by Schneider & Seitz (1995), we make the natural assumption that the intrinsic galaxy ellipticities $\varepsilon'$ are randomly oriented, so that the average intrinsic ellipticity $\langle\varepsilon'\rangle$ determined within each patch of the cluster field vanishes. The patches have to be chosen to be large enough to contain a sufficient number of galaxies for a reliable determination of the distortion and the magnification, but small enough such that the local lens properties do not change significantly within the patch. If $N'$ is the number of galaxies in such a patch, we require

$$\sum_{l=1}^{N'} \varepsilon'_l = 0 \, . \tag{5.1}$$

Using the inverse of Eq. (3.7), it can be shown (Schneider & Seitz 1995) that Eq. (5.1) translates to the requirement

$$\sum_{l=1}^{N'} \frac{\delta - \varepsilon_l}{1 - \mathrm{Re}(\delta \varepsilon_l^*)} = 0 \qquad (5.2)$$

for the distortion $\delta$ within the patch. Eq. (5.2) can be solved iteratively, yielding an estimator of $\delta$ within the given patch.

We choose the patches to be the cells of a rectangular grid, with $N_{\mathrm{grid}}$ cells on each side. All galaxy images closer than a smoothing radius $\Delta x$ from a grid point, and which have surface brightness within the specified surface-brightness bin, are included in the sum shown in Eq. (5.2). The smoothing radius $\Delta x$ can in principle be adapted to the strength of the signal, i.e., to the local average value of $|\varepsilon|$, as suggested by Seitz & Schneider (1994). In the context of this paper, however, this would introduce an unwanted complication into the problem. To see this, consider the "stack" of distortion fields corresponding to the individual surface brightness bins. The average distortions due to lensing increase towards lower surface brightness, because the average redshift increases. If we adapted the smoothing length to the strength of the signal, the smoothing length would decrease with decreasing surface brightness. Since smoothing reduces the signal amplitude, the comparison between the distortion fields in different surface-brightness bins would then be biased in the sense that the distortions for low-redshift galaxies would be underestimated compared to those of high-redshift galaxies. We could perhaps correct for this bias, but it is simpler to avoid the complication altogether by adopting the same fixed smoothing length $\Delta x$ for all the surface-brightness bins. We choose $\Delta x$ such that one smoothing disk contains on the order of ten galaxies. Since the galaxies are approximately uniformly distributed across the surface-brightness bins, we therefore require

$$\frac{n \pi \Delta x^2}{N_{\mathrm{bin}}} \approx 10 \;, \qquad (5.3)$$

or

$$\Delta x \approx 3 \sqrt{\frac{N_{\mathrm{bin}}}{n \pi}} \;. \qquad (5.4)$$

Following this procedure, we determine the distortion $\delta_{ijk}$ for each surface-brightness bin $i$, each patch $j$, and each cluster $k$, where $1 \leq i \leq N_{\mathrm{bin}}$, $1 \leq j \leq N_{\mathrm{grid}}^2$, and $1 \leq k \leq N_{\mathrm{clus}}$. We choose $N_{\mathrm{grid}} = 16$ in the following. Similarly, for each $ijk$, we calculate the local magnification parameter $r_{ijk}$ (cf. Eq. (2.2)) by comparing the average sizes $\langle R_{ijk} \rangle$ of the galaxy images in the local patch with the average sizes $\langle R_i' \rangle$ of galaxies of the same surface brightness in the control fields.

*5.1. Least-squares determination of redshifts and shear*

From Eq. (3.2) we see that a term independent of the source redshift can be factorized out of the expression for the convergence, and hence from the shear as well since it is linearly related to the convergence:

$$\kappa \equiv \kappa_\infty \, f(z_\mathrm{d}, z_\mathrm{s}) \quad , \quad \gamma \equiv \gamma_\infty \, f(z_\mathrm{d}, z_\mathrm{s}) \quad , \quad f(z_\mathrm{d}, z_\mathrm{s}) \equiv \frac{D_\mathrm{ds}}{D_\mathrm{s}} \;. \qquad (5.5)$$

The parameters $\kappa_\infty$, $\gamma_\infty$ refer to the convergence and shear for a source at "infinity," with $D_{\mathrm{ds}} = D_{\mathrm{s}}$. Let now a patch within a cluster lens be characterized by $\kappa_\infty$ and $\gamma_\infty$, and define

$$f_i \equiv \bar{f}(z_{\mathrm{d}}, z_i) , \tag{5.6}$$

where the bar denotes the average over all galaxies within surface-brightness bin $i$. Let us further normalize the $f_i$ by $f_1$. Then, we obtain from Eq. (5.5)

$$\kappa_i = \kappa_\infty \, f_i = \kappa_1 \, \frac{f_i}{f_1} \equiv \kappa_1 \, d_i \quad , \quad \gamma_i = \gamma_\infty \, f_i = \gamma_1 \, \frac{f_i}{f_1} \equiv \gamma_1 \, d_i . \tag{5.7}$$

Note the obvious point that if the mean redshift $\bar{z}_1$ of galaxies in the brightest surface-brightness bin ($i = 1$) has been measured spectroscopically as we assume, then $f_1$ can be calculated. Therefore, the lensing properties of each cluster are fully described either by knowing $(\kappa_\infty, \gamma_\infty)$ or $(\kappa_1, \gamma_1)$ across the cluster. In most of what follows, we make use of $(\kappa_1, \gamma_1)$.

Using Eq. (3.3) for the angular-diameter distances $D_{\mathrm{ds}}$ and $D_{\mathrm{s}}$, we can write

$$d_i = \frac{f_i}{f_1} = \frac{\zeta_1 - 1}{\zeta_1 - \zeta_{\mathrm{d}}} \left( 1 - \frac{\zeta_{\mathrm{d}} - 1}{\zeta_i - 1} \right) , \tag{5.8}$$

where we have used the abbreviations

$$\zeta_i \equiv \sqrt{1 + \bar{z}_i} \quad \text{and} \quad \zeta_{\mathrm{d}} \equiv \sqrt{1 + z_{\mathrm{d}}} \tag{5.9}$$

as in Eqs. (3.3) and (4.1). Note that all the numerically modeled clusters in our sample are at approximately the same redshift, hence the distance factors $d_i$ are the same for all clusters. If this were not the case, we would have had to introduce cluster-dependent distance factors $d_{ik}$, but the basic principle of the method described below would remain the same.

It is advantageous to use the quantity $g$ instead of the distortion $\delta$, because then the optimization procedure described below becomes linear. From Eq. (3.9),

$$g = \frac{\delta}{|\delta|^2} \left( 1 \pm \sqrt{1 - |\delta|^2} \right) , \tag{5.10}$$

where we must choose the sign to be opposite to that of the local Jacobian determinant $|\mathcal{A}|$ of the lens mapping. In general, this requires that the critical curves of the lens be known. For our purposes, however, we can safely adopt the negative sign in Eq. (5.10), because the strong smoothing inherent in the determination of $\delta$ guarantees that all of the cluster models to be reconstructed here are subcritical. We therefore convert the distortions $\delta_{ijk}$ into $g_{ijk}$ using only the negative sign in Eq. (5.10).

The overall goal of the algorithm is to use the $g_{ijk}$ and $r_{ijk}$ data to determine the distance factors $\{d_i\}$ corresponding to the surface-brightness bins $S_i$ and the convergence fields $\kappa_{1jk}$ and the shear fields $\gamma_{1jk}$ corresponding to surface brightness bin $S_1$ of the various lensing clusters. Once we have the $\{d_i\}$, it is straightforward to calculate the corresponding redshifts $\{\bar{z}_i\}$ through Eqs. (5.8) and (5.9). We break the calculation up into two nested iterations. In the inner loop, which is described in this subsection, we hold the $\kappa_{1jk}$ values fixed and determine the $\{d_i\}$ and $\gamma_{1jk}$ using the $g_{ijk}$ data. The outer loop, which is described in the next two subsections, updates the $\kappa_{1jk}$ values using the $r_{ijk}$ data. The overall structure of the algorithm is summarized in Sect. 5.5, and Fig. 5.

Given the $g_{ijk}$, we have to try and satisfy the relations (see Eq. (3.8))

$$g_{ijk} = \frac{\gamma_{ijk}}{1 - \kappa_{ijk}} = \frac{\gamma_{1jk} d_i}{1 - \kappa_{1jk} d_i} \tag{5.11}$$

simultaneously for all data points designated by the indices $(ijk)$. Correspondingly, we define the $\chi^2$ function

$$\chi_{\rm d}^2 \equiv \sum_{ijk} \frac{1}{2\sigma_{ijk}^2} \left[ g_{ijk}(1 - \kappa_{1jk} d_i) - \gamma_{1jk} d_i \right]^2 , \tag{5.12}$$

which we minimize with respect to the $d_i$ and $\gamma_{1jk}$, holding the $\kappa_{1jk}$ fixed at their current values. An estimate for the standard deviations $\sigma_{ijk}$ can be derived from the variances of the image ellipticities. For the purposes of our study, it is sufficient if the $\sigma_{ijk}$ correctly weight the relative reliability of the different "data points" $g_{ijk}$, because we shall not use the $\chi^2$ function itself to estimate the accuracy of the results. Rather, we compare the recovered results to the input data which we know, and we estimate the accuracy using Monte-Carlo techniques. If we were to apply these methods to real data, however, great care would have to be taken in estimating the $\sigma_{ijk}$.

Differentiating $\chi_{\rm d}^2$ with respect to $d_i$ and setting the result to zero yields

$$d_i = \left[ \sum_{jk} \frac{g_{ijk}}{\sigma_{ijk}^2} \left( g_{ijk} \kappa_{1jk} + \gamma_{1jk} \right) \right] \left[ \sum_{jk} \frac{1}{\sigma_{ijk}^2} \left( g_{ijk} \kappa_{1jk} + \gamma_{1jk} \right)^2 \right]^{-1} . \tag{5.13}$$

This equation gives the optimum estimate of the $d_i$ based on the given $g_{ijk}$. Of course, in order to calculate $d_i$, we need estimates of $\kappa_{1jk}$ and $\gamma_{1jk}$. As already mentioned, the determination of $\kappa_{1jk}$ is done through an outer iteration loop. In the first step of this outer loop, we set $\kappa_{1jk} = 0$. In later passes, $\kappa_{1jk}$ will have non-zero values. The shear factors $\gamma_{1jk}$ are however optimized in parallel with the $\{d_i\}$ by minimizing the same $\chi_{\rm d}^2$ given in Eq. (5.12). For the initial step, we assume the weak-lensing limit of Eq. (3.8) and obtain approximate estimates of $\gamma_{1jk}$ via Eq. (5.11); noting that $d_1 = 1$,

$$\gamma_{1jk} \approx g_{1jk}(1 - \kappa_{1jk}) . \tag{5.14}$$

These values of $\gamma_{1jk}$ are inserted together with the current values of $\kappa_{1jk}$ into the sums of Eq. (5.13) to determine $d_i$.

Once we have a first set of distance factors $d_i$, we can obtain improved estimates of the $\gamma_{1jk}$. Minimizing $\chi_{\rm d}^2$ of Eq. (5.12) with respect to $\gamma_{1jk}$, we obtain

$$\gamma_{1jk} = \left[ \sum_i \frac{g_{ijk} d_i}{\sigma_{ijk}^2} (1 - d_i \kappa_{1jk}) \right] \left( \sum_i \frac{d_i^2}{\sigma_{ijk}^2} \right)^{-1} , \tag{5.15}$$

which gives us a new set of $\gamma_{1jk}$ based on the current estimates of $\{d_i\}$ and $\kappa_{1jk}$. We thus use Eqs. (5.13) and (5.15) alternately, while keeping $\kappa_{1jk}$ fixed, to simultaneously optimize the $\{d_i\}$ and $\gamma_{1jk}$. This procedure usually converges within 5 to 10 iteration steps.

*5.2. Cluster reconstruction from shear*

Having determined the shear fields $\gamma_{1jk}$ for each cluster $k$ (or equivalently $\gamma_{\infty jk}$, see Eq. (5.7)), we can reconstruct the convergence fields $\kappa_{1jk}$ (or $\kappa_{\infty jk}$) using any of the reconstruction techniques described in the literature. For the present tests, we choose the simplest of these, namely the original method proposed by Kaiser & Squires (1993). (For other techniques, see Seitz & Schneider 1994, Kaiser et al. 1994, Bartelmann 1995b.) In this method, the convergence is related to the shear through the convolution

$$\kappa(\boldsymbol{x}) = -\frac{1}{\pi}\int_{\mathbb{R}^2} d^2 x' \, \mathrm{Re}\left[\mathcal{D}(\boldsymbol{x}-\boldsymbol{x}')\gamma^*(\boldsymbol{x}')\right] , \qquad (5.16)$$

where $\mathcal{D}(\boldsymbol{x})$ is the complex kernel

$$\mathcal{D}(\boldsymbol{x}) \equiv \frac{x_1^2 - x_2^2 + 2\mathrm{i}x_1 x_2}{x^4} . \qquad (5.17)$$

Given estimates of the shear $\gamma_{1jk}$ on a grid of points $\boldsymbol{x}_j$, we can approximate Eq. (5.17) by the sum

$$\kappa_{1jk} \approx -\frac{h^2}{2\pi}\sum_l \mathrm{Re}\left[\mathcal{D}(\boldsymbol{x}_j - \boldsymbol{x}_l)\gamma_{1lk}^*\right] , \qquad (5.18)$$

where $h$ is the separation of the grid points.

*5.3. Elimination of the mass sheet degeneracy*

The convergence obtained from applying Eq. (5.18) is not unique, but allows a transformation

$$\kappa \to 1 - \lambda + \lambda\kappa \qquad (5.19)$$

with arbitrary $\lambda \neq 0$ (e.g., Kaiser & Squires 1993, Schneider & Seitz 1995, Falco et al. 1985). For $\lambda \lesssim 1$, Eq. (5.19) corresponds to adding a sheet of constant surface mass density to the lens ($\lambda > 1$ corresponds to a sheet with negative mass density). Simultaneously, the shear is transformed according to $\gamma \to \lambda\gamma$, so that $g$ remains unchanged. What this means is that if the only information available is the image ellipticities, then the mass-sheet degeneracy (or $\lambda$ degeneracy) cannot be broken. As an aside we note that this statement is strictly valid only if the source galaxies are all located at a single redshift. For example, consider a situation where there are two source planes which are well separated in redshift. Then, the galaxies on the nearer plane feel convergence and shear $\kappa_1$ and $\gamma_1$, while those on the farther plane feel $\kappa_2$ and $\gamma_2$, with

$$(\kappa_2, \gamma_2) = (\kappa_1, \gamma_1)d_2 .$$

If we rescale $\kappa_1$ and $\gamma_1$ according to the transformation (5.19), we obtain for $\kappa_2$ and $\gamma_2$

$$\kappa_2 = (1 - \lambda + \lambda\kappa_1)d_2 \quad , \quad \gamma_2 = \lambda\gamma_1 d_2 .$$

Hence the sources on plane 2 are deformed by

$$g_2 = \frac{\gamma_2}{1 - \kappa_2} = \frac{\lambda\gamma_1 d_2}{1 - (1 - \lambda + \lambda\kappa_2)d_2} ,$$

420                                                                  *The Lens Parallax Method:*which is not independent of $\lambda$. Therefore, the mass-sheet degeneracy can in principle be broken using just image ellipticities, provided the sources are distributed over a sufficiently wide range of redshift. However, this method would probably be very inaccurate in practice.

Our method to break the mass-sheet degeneracy makes use of measurements of the inverse lens magnifications $r_{ijk}$. The magnification of the lens is given by

$$\frac{1}{r} = \mu \equiv \frac{1}{\det \mathcal{A}} = \frac{1}{(1-\kappa)^2 - |\gamma|^2} \; . \quad (5.20)$$

Since this expression is not invariant under the $\lambda$ transformation, measurements of the magnification can in principle be used to determine $\lambda$. The calculation proceeds as follows.

For each $ijk$, we have the relation

$$r_{ijk} \equiv \left(\frac{\langle R'_i \rangle}{\langle R_{ijk} \rangle}\right)^2 = \frac{1}{\mu_{ijk}} = [1 - (1 - \lambda_k + \lambda_k \kappa_{1jk})d_i]^2 - (\lambda_k \gamma_{1jk} d_i)^2 \; , \quad (5.21)$$

which must be satisfied to within the noise. This equation can be used to define a $\chi^2$ function as before, but now for each cluster separately:

$$\chi^2_{\lambda,k} = \sum_{ij} \frac{1}{2\sigma^2_{ij}} \left\{ r_{ijk} + (\lambda_k \gamma_{1jk} d_i)^2 - [1 - (1 - \lambda_k + \lambda_k \kappa_{1jk})d_i]^2 \right\}^2 \; . \quad (5.22)$$

We minimize $\chi^2_{\lambda,k}$ with respect to the $\lambda_k$ to determine the optimum mass sheet. The result is a global scale factor $\lambda_k$ for each cluster, determined such as to maximize the overlap between the de-magnified sizes of the galaxies in the cluster fields with those of the galaxy sizes in the control fields. In performing this minimization, however, we need to allow for the fact that only those galaxies may be included in the determination of $\langle R_{ijk} \rangle$ which after de-magnification still fulfil the detection criterion of Eq. (4.15). This is to avoid any magnification bias in the calculations. Since the de-magnification depends on the $\lambda_k$, the average sizes $\langle R_{ijk} \rangle$ have to be determined anew each time $\lambda_k$ is changed during the minimization procedure.

Having determined the best-fit set of $\lambda_k$, we can rescale the convergence fields $\kappa_{1jk}$ according to Eq. (5.19) and re-optimize the $d_i$ and $\gamma_{1jk}$ by the method described in Sect. 5.1. This iteration between an outer loop, where the $\lambda_k$ and $\kappa_{1jk}$ are determined, and an inner loop, where $\{d_i\}$ and $\gamma_{1jk}$ are optimized, is continued until convergence is achieved in both loops.

Note that in this method we use the magnifications to determine only one parameter per cluster, namely the mass sheet parameter $\lambda_k$, whereas the *shape* of the convergence field $\kappa_{1jk}$ is obtained entirely from the measurements of image distortions via the Kaiser & Squires method described in Sect. 5.2. The reason for this approach is that shear measurements are normally expected to be more accurate than magnifications. Since we have chosen the smoothing radius $\Delta x$ such that there are $N' \approx 10$ galaxies in a smoothing disk, and since the shear is proportional to the galaxy ellipticity in the weak-lensing limit (cf. Eq. (2.1)), the error in the estimate of $\gamma_{1jk}$ in a given patch is

*Method of analysis*  21

$$\Delta\gamma \sim \frac{\sigma_\varepsilon}{\sqrt{2\,N'\,N_{\rm bin}}} \sim 0.02\;,$$

where the factor of two in the denominator accounts for there being two components of the shear which can be measured independently, and we have accounted for the fact that the data on all $N_{\rm bin} = 10$ surface-brightness bins contribute to the estimate of $\gamma_{1jk}$. On the other hand, we have from Eq. (2.2) that the error in the estimate of the local magnification is

$$\Delta\kappa \sim \frac{1}{\sqrt{N'\,N_{\rm bin}}} \frac{\Delta R}{R} = \frac{\Delta\ln(R)}{\sqrt{N'\,N_{\rm bin}}} \sim 0.05\;,$$

where we have used $\Delta\ln(R) \sim 0.5$ as indicated by our simulated galaxy population. Thus we have

$$\Delta\kappa \sim 2.5\Delta\gamma\;,$$

so that it is preferable to use the shear estimates rather than the magnifications. However, it may be possible in some cases to determine $\kappa_{1jk}$ directly from the magnification estimates $r_{ijk}$ in each patch. This is described in the next subsection.

*5.4. Cluster mass reconstruction directly from galaxy sizes*

In this alternative method we use Eq. (5.21) directly to find a least-squares estimate of the convergence field. The $\lambda_k$ had to be introduced in the previous subsection only because a cluster reconstruction based on image distortions alone allows for transformations of $\kappa$ according to Eq. (5.19). If we attempt to infer the convergence directly from the galaxy sizes, the $\lambda_k$ can be set to unity in Eq. (5.21). Then, we can write

$$r_{ijk} = (1 - \kappa_{1jk}d_i)^2 - \gamma_{1jk}^2 d_i^2\;, \tag{5.23}$$

and the $\kappa_{1jk}$ can be derived by minimizing the $\chi^2$ function

$$\chi_{\rm r}^2 = \sum_{ijk} \frac{1}{2\sigma_{ijk}^2} \left[\kappa_{1jk}d_i - 1 + \left(r_{ijk} + \gamma_{1jk}^2 d_i^2\right)^{1/2}\right]^2\;. \tag{5.24}$$

The minimization yields

$$\kappa_{1jk} = \left(\sum_i \frac{d_i}{\sigma_{ijk}^2} \left[1 - \left(r_{ijk} + \gamma_{1jk}^2 d_i^2\right)^{1/2}\right]\right) \left(\sum_i \frac{d_i^2}{\sigma_{ijk}^2}\right)^{-1}\;. \tag{5.25}$$

We discuss the relative merits of this and the previous method of estimating $\kappa_{1jk}$ when we describe the results. Note, however, that these two methods represent two limits of how to use the galaxy size information. It is possible to develop intermediate schemes where the magnification measurements are used in an optimal way, depending on the relative accuracies of the measured shears and magnifications. We do not investigate such refinements in this paper.

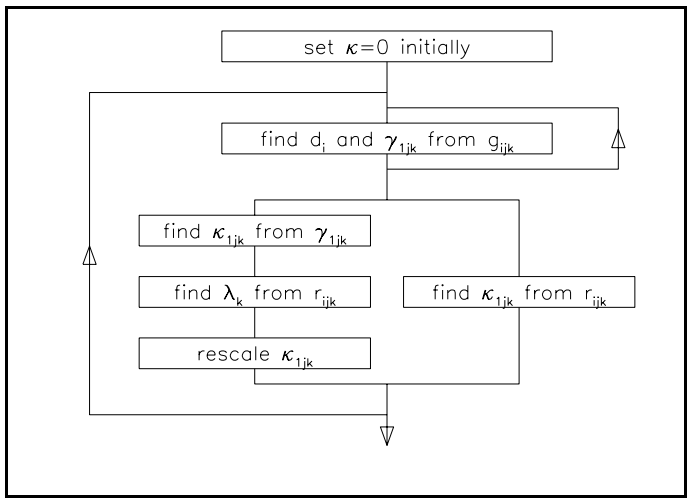

Figure 5.– Flowchart summarizing the algorithm described in the text.

### 5.5. Summary of the algorithm

The iterative algorithm described above proceeds as follows, as summarized in the flow diagram of Fig. 5.

Initially, the convergence field $\kappa_{1jk}$ is set to zero everywhere, corresponding to the weak-lensing limit. The measured image distortions in the gridded patches of the clusters, $g_{ijk}$, are used to solve iteratively for the best-fit set of distance factors $d_i$ and shear values $\gamma_{1jk}$. The $d_i$ are related to the mean redshifts $\bar{z}_i$ of the galaxies in the surface-brightness bins through

$$d_i \equiv \frac{\zeta_1 - 1}{\zeta_1 - \zeta_d}\left(1 - \frac{\zeta_d - 1}{\zeta_i - 1}\right), \qquad (5.26)$$

where $\zeta_d \equiv \sqrt{1+z_d}$ and $\zeta_i \equiv \sqrt{1+\bar{z}_i}$. Assuming the average redshift $\bar{z}_1$ of the galaxies in the brightest surface-brightest bin ($i = 1$) is known, the distance factors $d_i$ can be converted to redshifts $\bar{z}_i$.

Once convergence is achieved in the inner loop, we have two choices. In one choice, we determine the convergence fields $\kappa_{1jk}$ of the clusters using shear fields $\gamma_{1jk}$ and one of the standard reconstruction techniques, such as that due to Kaiser & Squires (1993). The reconstruction from the shear field requires a subsequent calibration of the global scale parameter $\lambda_k$, which is done such that the average galaxy sizes $\langle R_{ijk}\rangle$ are consistent with the sizes $\langle R'_i\rangle$ in the control fields. In the second approach, we solve directly for $\kappa_{1jk}$ using the galaxy sizes in each patch of the sky and calculating the local inverse magnification $r_{ijk}$. In either approach, once the $\kappa_{1jk}$ are obtained, these values are inserted in place of $\kappa = 0$ into the inner loop and the $d_i$ and $\gamma_{1jk}$ are optimized once again. Using these, a new set of $\kappa_{1jk}$ are again calculated, and so on. The algorithm is continued until convergence is achieved, which usually takes three to five iterations in the outer loop and five to ten iterations each time in the inner loop.

Figure 6 illustrates the convergence of the distance factors towards their final values in the course of four iteration steps in the outer loop of the flow diagram of Fig. 5. The heavy line shows the true distance factors, and the solid line shows the recovered distance factors after convergence was achieved. The dotted lines display the intermediate results after one, two, and three iteration steps, and the arrow indicates the direction of progression of the values. The results obtained in the first iteration step substantially overestimate the true values of $d_i$ because of the initial assumption that $\kappa = 0$. As $\kappa$ is built up during successive iterations, the recovered distance factors gradually decrease until the algorithm converges with a fairly accurate reconstruction of the distances.

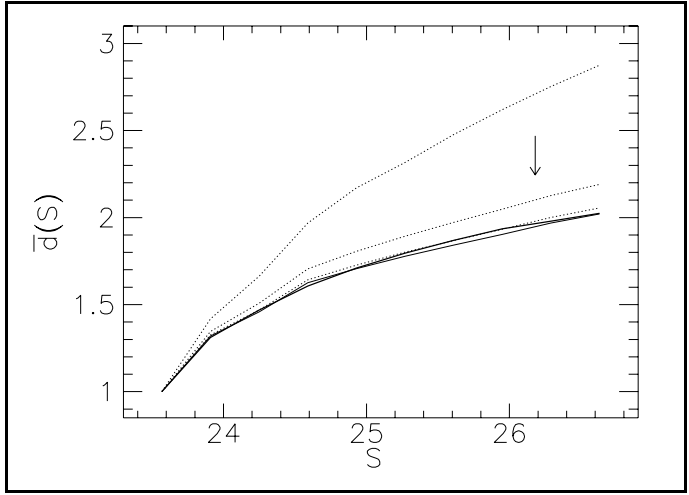

Figure 6.– Example showing the convergence of the distance factors $\bar{d}_i$ towards their final value in successive passes through the outer loop of the algorithm. The heavy line shows the true distance factors as a function of apparent surface brightness. The dotted lines show intermediate results obtained during the iteration, and the solid line shows the result to which the algorithm finally converged. The arrow indicates the progression of the iteration.

## 6. Results

### 6.1. The redshift distribution of the source galaxies

Figure 7 shows two examples of the recovered average distance factors $\bar{d}_i$ of the source galaxies as a function of surface brightness $S_i$, derived using the algorithm described above. The frame on the left shows the results obtained with the full sample of twelve cluster fields, while the frame on the right corresponds to a reduced sample of six clusters.

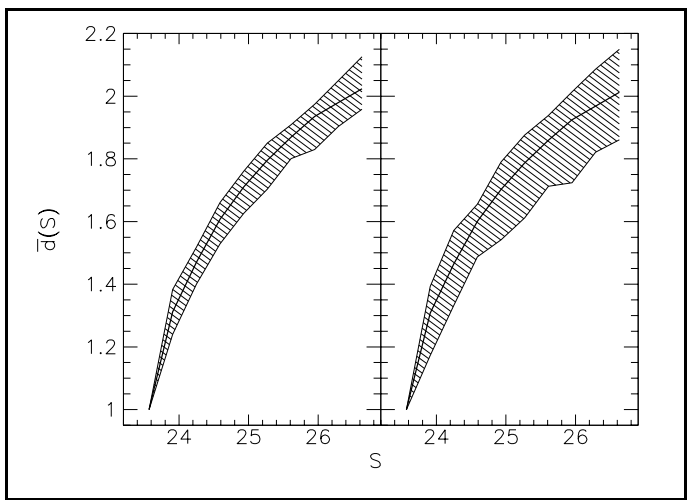

Figure 7.– Average galaxy distance factors $\bar{d}_i$ as a function of apparent surface brightness. The heavy lines display the true distance factors, while the reconstructed results are shown by the hatched stripes. The widths of these stripes show the 1-$\sigma$ deviations obtained from ten independent simulation runs, either with twelve (left panel) or six (right panel) clusters each. We see that the distance factors are recovered quite accurately.

The widths of the hatched stripes in the figure show the standard (1-$\sigma$) deviation of the distance factors obtained from ten independent simulation runs. The heavy lines display the true distance factors for comparison. These results clearly demonstrate that our algorithm recovers the distance factors of the galaxies quite reliably, with the 1-$\sigma$ deviation being only $\approx 0.07$ at the high-redshift end for the sample of 12 clusters. Reducing the sample size to 6 clusters approximately doubles the error bars.

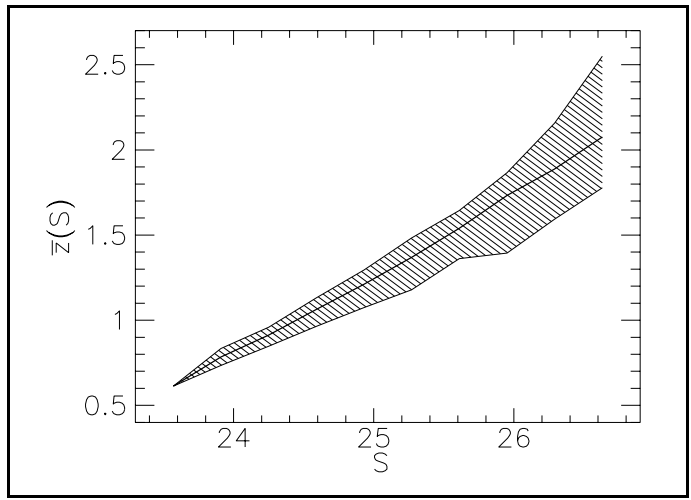

Figure 8.– Average galaxy redshifts $\bar{z}_i$ as a function of apparent surface brightness. The figure shows the same data as in the left panel of Fig. 6, but the distance factors have been converted to redshifts. The uncertainty in the redshifts is larger than that in the distance factors at low surface brightness, because the distance factors depend only weakly on redshift at high $z$.

Figure 8 shows the mean redshifts reconstructed by the algorithm, where the $d_i$ have been converted to $\bar{z}_i$ through Eq. (5.26). Given the error bars $\Delta d_i$ of the distance factors, the error bars on the redshifts are determined by

$$\Delta z_i = \left(\frac{\mathrm{d}\,d(z)}{\mathrm{d}\,z}\right)^{-1}\bigg|_{z=z_i} \Delta d_i \ . \tag{6.1}$$

For increasing redshift, $d(z)$ becomes flat, and therefore $\Delta z$ grows more rapidly with increasing $z$ than $\Delta d$. This reflects the fact that gravitational lenses can only reliably distinguish between the redshifts of sources which are not too far behind the lenses. Fig. 8 shows that, with a sample of twelve rich clusters at moderate redshifts ($z_d \sim 0.35$), we can hope to derive faint galaxy redshifts with an accuracy of $\Delta z \approx 0.1 - 0.2$ for galaxies at redshifts between $z \approx 1 - 1.7$, and $\Delta z \approx 0.3$ at $z \approx 2$.

In order to estimate the influence of the intrinsic width $\sigma_\mathrm{I}$ of the surface-brightness distribution $p_\mathrm{I}$ of Eq. (4.5), we have repeated the calculations with $\sigma_\mathrm{I}$ doubled to $1.0 I_0$. We find that the 1-$\sigma$ error bars increase from $\approx 5\%$ to $\approx 8\%$ at the high-redshift end of $\bar{d}(S)$. This means that the results are not unduly dependent on the particular assumptions we made regarding the narrowness of the surface brightness distribution of galaxies. Thus it appears that we can significantly relax the tightness of the intrinsic surface-brightness distribution without altering the conclusions very much. The results are more sensitive to the width $\sigma_\varepsilon$ of the intrinsic ellipticity distribution. If we increase $\sigma_\varepsilon$ from 0.3 to 0.5, the recovered distance factors $d_i$ are systematically too small and reach only $\approx 90\%$ of their true values for the faintest galaxies. This bias is a consequence of the fact that the ellipticities $\varepsilon$ are bounded by unity from above, and the resulting non-linearity leads to an attenuation of the signal $\delta_{ijk}$ when it is estimated by the methods described here. The bias can presumably be removed by developing better methods to estimate $\delta_{ijk}$, but the reduced accuracy due to a larger spread of $\sigma_\varepsilon$ is unavoidable.

## 6.2. Accuracy of the mass reconstruction

The second aspect of our algorithm is that it allows us to construct convergence maps $\kappa_{1jk}$ of the lensing clusters from which mass maps $\Sigma(\boldsymbol{x})$ are obtained via Eq. (3.2). As explained earlier, the maps of $\kappa_{1jk}$ are either directly derived from the galaxy sizes (Sect. 5.4), or they are reconstructed from the shear fields (Sect. 5.2) and scaled according to Eq. (5.19) such that the scale factors $\lambda_k$ optimize the agreement between the de-magnified sizes of the galaxy images in the cluster fields with those in the control fields (Sect. 5.3). As a global measure of the accuracy of the mass reconstructions, we determine the ratio $F(r)$ between the reconstructed mass and the true mass enclosed within radius $r$ around the cluster center,

$$F(r) \equiv \frac{M_{\rm rec}(r)}{M_{\rm true}(r)}, \qquad (6.2)$$

where $M_{\rm rec,true}(r)$ are the reconstructed and true cluster masses, respectively. In Fig. 9, we show the cumulative function $F(r)$ and its differential version,

$$f(r) \equiv \left(\frac{{\rm d}M_{\rm rec}(r)}{{\rm d}r}\right) \left(\frac{{\rm d}M_{\rm true}(r)}{{\rm d}r}\right)^{-1}, \qquad (6.3)$$

for the two reconstruction methods described. In practice, we replace the derivative in (6.3) by the masses inside rings of finite width.

Focusing first on the upper left frame of the figure, we see that the determination of $\kappa_{1jk}$ from galaxy sizes alone very accurately reproduces the true total cluster mass contained within radius $r$, once $r$ becomes larger than $\approx 1'$. Interior to $1'$, the reconstructed mass fraction drops towards $F(r) \approx 0.7$. This is a consequence of the smoothing, which broadens mass peaks and therefore lowers the contrast, shifting mass outwards from the central peak. In the differential version of the results, displayed in the upper right frame, it is evident that the mass is overestimated at intermediate radii $r \approx 1'$. This is a result of the algorithm's attempt to reconstruct accurately the average magnification in the field. The method thus compensates for the mass deficit towards the cluster center caused by smoothing, by adding more mass at slightly larger radii.

For the mass reconstruction from the shear fields, with $\lambda$ scaling, the behaviour of the reconstructed mass fractions $f(r)$ and $F(r)$ is somewhat different. The differential function approaches unity for $r \gtrsim 2'$ and decreases steadily towards $f(r) \approx 0.45$ at the cluster center. The cumulative function $F(r)$ therefore stays below unity across the whole field, approaching $F(r) \approx 0.85$ towards the field boundary. Once again, smoothing is part of the explanation, since it broadens mass peaks and re-distributes mass over a larger area. However, in addition, these curves reflect a systematic effect inherent in the original version of the Kaiser & Squires reconstruction employed here. Eq. (5.18) shows that the method requires an integration to be performed over the entire real plane, whereas in reality the integration is confined to the finite field containing the data. This causes the method to reconstruct a mass distribution with vanishing total mass within the field, which it does by introducing spurious negative troughs towards the field boundaries. In the corners of the field, however, the reconstructed surface-mass density increases again to spurious positive values. The determination of the $\lambda_k$ scale factors using galaxy sizes attempts to correct for this effect by adjusting the average mass density in the field so as to fit the average observed magnification. However, since we do the mass comparison on circles, the regions of positive mass density in the corners are excluded from $f(r)$ and $F(r)$, and part of the mass which accounts for the overall mass balance is neglected. Therefore, the measures employed here discriminate against the Kaiser & Squires method, but they do so because of a systematic error in the method. More elaborate variants of the Kaiser & Squires method have been suggested which are designed to avoid this systematic effect (Schneider 1995, Kaiser et al. 1994, Bartelmann 1995b). We have not yet tested these methods in conjunction with our algorithm.

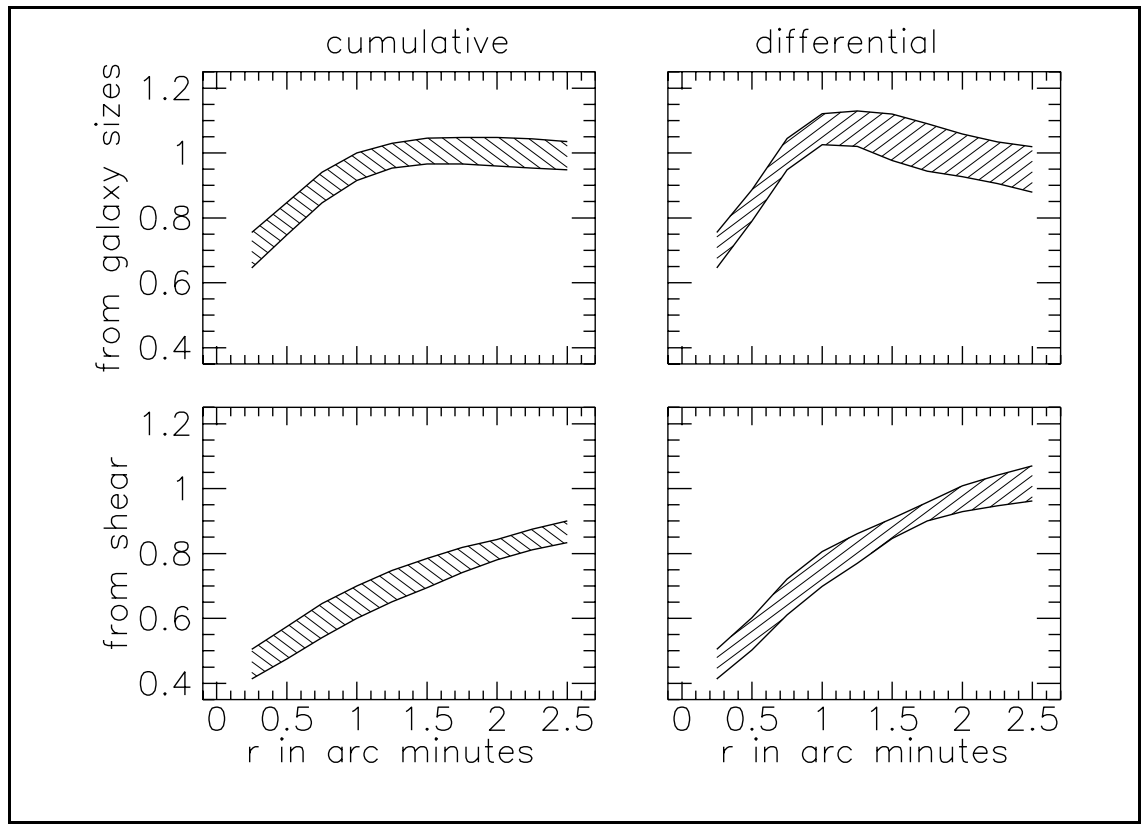

Figure 9.– Ratio $F(r)$ between the reconstructed mass and the true mass of the lens within radius $r$ ("cumulative", left frames) and its differential version $f(r)$ (right frames; see Eqs. (6.2) and (6.3)). The top panels show the results from reconstructing surface-mass densities from galaxy sizes. The results displayed in the bottom panels were obtained from the shear data via the Kaiser & Squires reconstruction method and subsequently rescaled to fit the observed magnifications.

Finally, we show in Fig. 10 surface-mass density maps of one cluster of our sample. The upper left frame shows the contours of the original cluster model, and the upper right frame shows the original model after it has been smoothed by the same smoothing

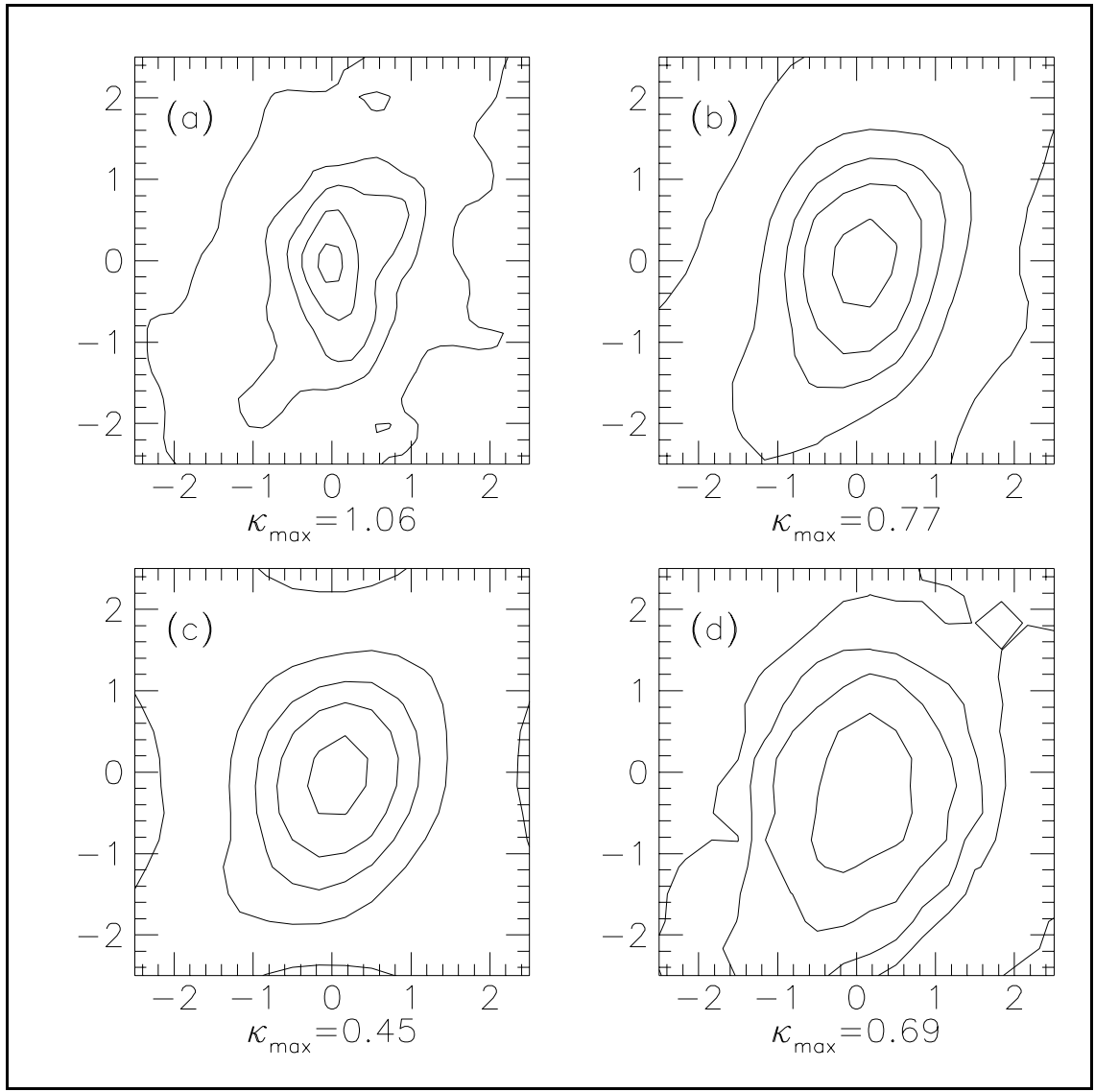

Figure 10.– Surface-mass density maps of one of the simulated clusters. Frame (a): original cluster model; frame (b): original model smoothed to the same degree as the reconstructions; frame (c): result of applying the Kaiser & Squires reconstruction technique, as described in Sect. 5.2 and 5.3; frame (d): reconstruction using only galaxy sizes, as described in Sect. 5.4. The scale is given in arc minutes, and the peak values of $\kappa$ are given below the images. The contours have been drawn at $\{10, 33, 50, 67, 90\}\%$ of the difference between the peak and the surroundings.

length as in the reconstructions. The lower left frame shows the map obtained from the Kaiser & Squires reconstruction (Sects. 5.2 and 5.3), and the lower right frame displays the map derived from galaxy sizes alone (Sect. 5.4). Note that the peak values, which are given below the frames, differ quite significantly from frame to frame. The contour levels have been chosen to represent $\{10, 33, 50, 67, 90\}\%$ of the difference between the peaks and the surroundings. Clearly, the mass map obtained from the Kaiser & Squires method has less noise than the map obtained using galaxy sizes alone. The Kaiser &

Squires method is therefore superior for obtaining a detailed reconstruction of the shape of the cluster, especially close to the cluster center. However the method performs less well for estimating global quantities such as the total mass in the cluster for which the galaxy-size approach is superior. It seems clear that one should be able to develop hybrid schemes which make use of the strengths of both methods in an optimal way. We leave the discussion of such techniques to a future paper.

## 7. Summary and discussion

We propose to call the technique described in this paper to determine relative distances to faint galaxies the *lens parallax* method. Although the technique does not employ true trigonometric parallaxes, it is nevertheless a purely geometrical method. In essence, a given local patch of a lensing cluster produces a differential bending of light rays which is described by the local values of the convergence $\kappa_\infty$ and the shear $\gamma_\infty$, the latter being a complex number with two components. (Equivalently, as we have done in this paper, we could describe the lens properties in terms of $\kappa_{1jk}$ and $\gamma_{1jk}$ corresponding to the brightest surface brightness bin.) The ray deflections distort and magnify the images of background sources in such a way that the magnitude of the effects depends through simple geometry on the distance between the lens and the source. The analogy with trigonometric parallax is fairly close.

The paper is based on two key ideas. First, we show how the lens parallax method can be used to determine the redshifts of faint galaxies as a function of surface brightness. Related ideas have been discussed before by Smail et al. (1994) and especially Kaiser et al. (1994) who showed how it is possible in principle to use image distortions to estimate redshifts as a function of galaxy magnitude. A new feature of our approach is that we label galaxies by means of their surface brightness rather than magnitude. Because surface brightness is conserved under the action of gravitational lensing, it is a convenient way of comparing galaxies in lensed fields with those in control fields. Furthermore, based on our knowledge of galaxy properties, it appears that the surface brightness is likely to be more tightly correlated with galaxy redshift than magnitudes (see Figs. 1 and 2). This again makes surface brightness a convenient galaxy label for our technique. The second idea in this paper is that we suggest comparing galaxy angular sizes in lensed fields with the sizes of galaxies in unlensed control fields, thereby measuring the local magnification. Once again, the comparison is done by using surface brightness as a galaxy label. By obtaining an estimate of the magnification, even if it is relatively inaccurate, we show that it is possible to eliminate the mass sheet degeneracy (Falco et al. 1985, Kaiser & Squires 1993, Schneider & Seitz 1995) which has plagued previous methods of determining cluster masses through weak lensing. Broadhurst et al. (1995) suggested a closely related idea, where they compare galaxy counts in the lensed fields against the counts in blank control fields to determine the magnification. We suspect that our method, which directly compares angular sizes of galaxies, is simpler and less model-dependent.

We have described and tested a global iterative algorithm which allows us simultaneously to reconstruct the redshift distribution of the faint blue galaxies as a function of surface brightness, and the surface mass distributions of the lensing galaxy clusters. The method requires the ellipticities, sizes, and surface brightnesses of galaxy images to be measured in a number of cluster fields and "empty" control fields. In addition, the cluster redshifts and the redshifts of the brightest background galaxies need to be known, the latter in order to calibrate the relative distance scale which the method provides. The numerical simulations we have performed show that the algorithm is feasible. Observations of $\sim 2000$ faint galaxies each in $\approx 10$ cluster fields and an equal number of control fields should be sufficient to achieve a relative accuracy of $\approx 10-20\%$ in the galaxy redshifts. The total masses derived for the cluster lenses are accurate to $\approx \pm 5\%$.

We must emphasize that the lens parallax method is quite model-independent and does not rely on any specific assumptions about the properties of distant galaxies or their evolution. All we require is that there should be a reasonably tight relation between surface brightness and redshift. This is expected on theoretical grounds since the apparent surface brightness of a galaxy is a steep function of redshift, e.g. $I \propto (1+z)^{\alpha-3}$ for a spectral index $\alpha$ (Eq. 4.7). There is also direct observational evidence for a correlation (Koo & Kron 1992, Tyson 1994). Although we had to invent a specific model of intrinsic galaxy properties in order to test the method, the only essential feature of this model for our purposes is that the surface brightness depends steeply on redshift. Other than this, we make no serious assumptions. In particular, we do not require that the galaxies at high redshift should necessarily be similar to local galaxies or that their angular sizes should evolve in any particular way (for instance, even if galaxy angular sizes turn out to be independent of $z$ the method will still work). The reason is that the control fields allow us to calibrate all of the necessary galaxy properties as a function of surface brightness, and our method makes use of only differential effects between the lensed and unlensed fields. Indeed, because the method is so model-independent, it appears to be a powerful method for studying high redshift galaxies. Once we obtain $\bar{z}_i$ as a function of surface brightness $S_i$, we can expect to have quite a lot of information on the luminosities, number densities, colors, etc. of galaxies at high redshift. This information will provide significant constraints on models of galaxy formation and evolution at high $z$.

The second aspect of our method, namely the determination of accurate mass distributions of the lensing clusters, seems equally promising. Masses derived from gravitational lensing have the advantage that they make no assumptions regarding the virial state of the cluster or the distribution of mass versus light. Unfortunately, mass estimates from weak lensing have tended to be somewhat uncertain because of the mass sheet degeneracy, while mass estimates from giant arcs are usually not very well constrained (Bartelmann 1995a). Our method appears to be able to derive quite accurate and unambiguous enclosed masses of clusters and even fairly good two-dimensional mass maps. The algorithm we have used for the tests presented in this paper is fairly basic. It is conceivable that by optimizing the method to make the best use of the information on image distortions and magnifications we may be able to achieve even higher accuracy.

The methods described here can be extended in several ways. First, instead of using only surface brightness to characterize the galaxy population, we could also include other properties that are invariant to lensing, e.g. colors (assuming there is no reddening due to the lenses). We might thus be able to obtain the mean redshift of galaxies as a function of both brightness and color, thus achieving a more detailed description of galaxy evolution. Another possibility is that by comparing the results from lenses at different redshifts $z_\mathrm{d}$, we can test for the internal consistency of the independent distance ladders $\{d_i\}$ obtained from each cluster. In principle, this will allow us to constrain the model of the universe. Gravitational lenses are particularly effective at distinguishing models with a large cosmological constant $\Lambda$ (Turner 1990, Kochanek 1993). The lens parallax method may conceivably provide a purely geometrical technique for constraining the value of $\Lambda$. Finally, we can test whether the entire lens effect is due only to the action of the cluster, as assumed in our analysis, or whether there are additional lenses in the line of sight at different redshifts. For instance, if there is only one lens in the field, then the shear angle $\varphi_\gamma(S)$ in a given patch of the lens will be independent of apparent surface brightness $S$. If the data show instead a variation of $\varphi_\gamma$ with $S$ (or equivalently source redshift), then we could infer that there are significant additional shear contributions at redshifts $z > z_\mathrm{d}$ which introduce differential effects as a function of source redshift. This could potentially be a powerful diagnostic of structure at high redshift.

The role of systematic effects in the lens parallax method is unclear at the moment. The method requires that we be able to measure the ellipticities, scale lengths and surface brightnesses of very faint galaxies with good accuracy. The effect of atmospheric seeing may be particularly serious, and it is possible that the method may require space-based observations in order to avoid errors due to variable seeing. Even with space observations, there may be effects due to the finite pixel size of the imaging camera which we have not included in our analysis.

*Acknowledgements.*We thank Paul Schechter for valuable advice on modeling the intrinsic properties of faint galaxies, and Nick Kaiser, Chris Kochanek, and Peter Schneider for comments and suggestions. Matthias Steinmetz and Achim Weiss contributed substantially to the numerical cluster simulations which we have used here. This work was supported in part by NSF grant AST 9423209 and by the Sonderforschungsbereich SFB 375-95 of the Deutsche Forschungsgemeinschaft.